\documentclass[aps,prl,reprint,showpacs,superscriptaddress,longbibliography]{revtex4-1}
\usepackage[dvipdfmx]{graphicx}

\usepackage{pdfpages} % include pdfs
\usepackage{pgffor} % for loops
\makeatletter
\AtBeginDocument{\let\LS@rot\@undefined}
\makeatother

\def\supplementfilename{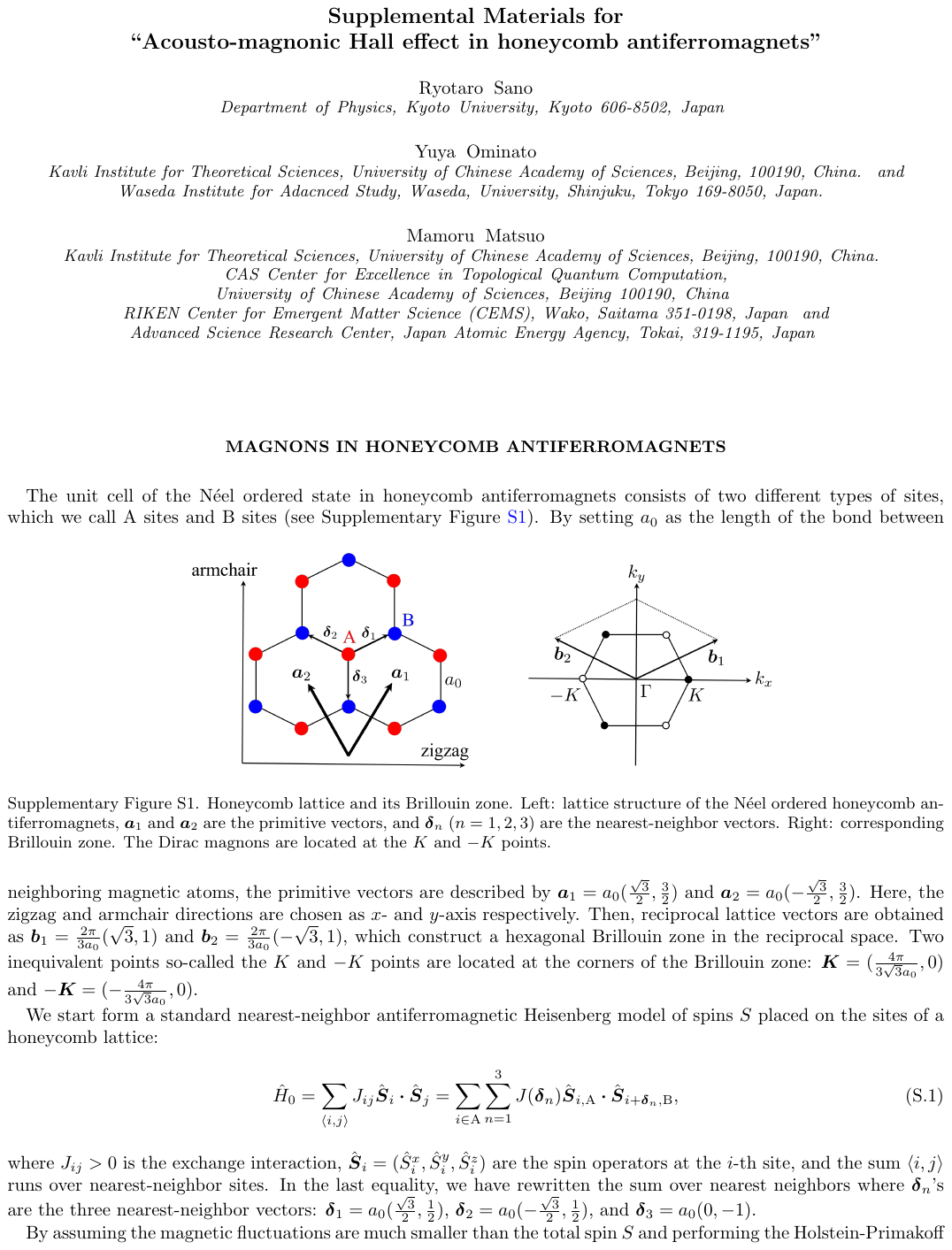}

\pdfximage{\supplementfilename}
\def\numbersupplementpages{\the\pdflastximagepages}

\newif\ifarXiv
\arXivtrue 
\usepackage{amsmath,amssymb,amsfonts}
\usepackage{comment}
\usepackage{color}
\usepackage[T1]{fontenc}
\usepackage{textcomp}
\usepackage{hyperref}
\hypersetup{
setpagesize=false,
colorlinks=true,
linkcolor=blue,
citecolor=red,
}
\usepackage{bm}
\usepackage{mathrsfs}
\usepackage{physics}
\usepackage{svg}
\usepackage{ulem}
\usepackage{multirow}

\makeatletter
\let\MYcaption\@makecaption
\makeatother
\usepackage{subcaption}
\makeatletter
\let\@makecaption\MYcaption
\makeatother

\begin{document}

\title{Acousto-magnonic spin Hall effect in honeycomb antiferromagnets}%

\author{Ryotaro Sano}
\email{sano.ryotaro.52v@st.kyoto-u.ac.jp}
\affiliation{%
Department of Physics, Kyoto University, Kyoto 606-8502, Japan
}%

\author{Yuya Ominato}
%\email{}
\affiliation{%
Kavli Institute for Theoretical Sciences, University of Chinese Academy of Sciences, Beijing, 100190, China.
}%

\affiliation{%
Waseda Institute for Advanced Study, Waseda University, Shinjuku, Tokyo 169-8050, Japan.
}%

\author{Mamoru Matsuo}

\affiliation{%
Kavli Institute for Theoretical Sciences, University of Chinese Academy of Sciences, Beijing, 100190, China.
}%

\affiliation{%
CAS Center for Excellence in Topological Quantum Computation, University of Chinese Academy of Sciences, Beijing 100190, China
}%
\affiliation{%
RIKEN Center for Emergent Matter Science (CEMS), Wako, Saitama 351-0198, Japan
}%
\affiliation{%
Advanced Science Research Center, Japan Atomic Energy Agency, Tokai, 319-1195, Japan
}%

\date{\today}

\begin{abstract}
The recently discovered van der Waals antiferromagnets have suffered from the lack of a comprehensive method to study their magnetic properties. Here, we propose an AC intrinsic magnon spin Hall current driven by surface acoustic waves as a novel probe for such antiferromagnets. Our results pave the way towards mechanical detection and manipulation of the magnetic order in two-dimensional antiferromagnets. Furthermore, they will overcome the difficulties with weak magnetic responses inherent in the use of antiferromagnets and hence provide a building block for future antiferromagnetic spintronics.
\end{abstract}

\maketitle
{\it Introduction.---}
Intrinsic magnetism in two-dimensional (2D) materials has long been sought-after but believed to hardly survive due to the enhanced thermal fluctuations according to the Mermin-Wagner theorem~\cite{Mermin-Wagner1966PRL}. However, the recent discovery of mechanically exfoliated van der Waals (vdW) magnets~\cite{Gong2017Nat,Huang2017Nature} has revealed that the magnetic anisotropy can resist the thermal agitation and stabilize long-range magnetic order in the 2D limit at finite temperatures~\cite{Burch2018Nature,Gong2019Science,Gibertini2019NatNanotech,Mak2019NatRevPhys,Huang2020NatMater,Cortie2020AdvFuncMat,Zhang2021Nanoscale,Och2021Nanoscale,Jiang2021APR,Xu2022Microstructures,Wang2022ACSNano,Kurebayashi2022NatRevPhys,Huang2018Natnanotech,Jiang2018NatMater,Jiang2018Natnanotech,Song2018Science,Yang2019APL,Song2019NatMater,Chen2019Science,Jin2020NatMater,Fei2018NatMater,Deng2018Nature,Klein2018science,Wang2019NatNanotech,Zhang2019Nanolett,Kim2019natelectronics,Purbawati2020ACSAppl}. Especially, transition metal phosphorus trichalcogenides $M$P$X_3$ ($M=$ Mn, Fe, Ni; $X=$ S, Se) are a family of vdW antiferromagnets, and are easily exfoliatable down to the monolayer limit due to their
weak vdW interlayer interaction~\cite{Du2016acsnano}. These materials share the same honeycomb lattice structure but the bulk antiferromagnetic (AFM) phase at low temperatures varies depending on the magnetic elements~\cite{JoyP1992PRB,Ressouche2010PRB,Wildes2015PRB,lancon2018PRB,Long2020NanoLett}. It therefore provides an ideal platform to investigate magnetism and magnetic excitations in the 2D limit. Furthermore, compared to ferromagnets, antiferromagnets exhibit ultrafast dynamics in the terahertz regime, null stray field, and robustness against the external magnetic field perturbation~\cite{Jungwirth2016NatNanotech,Baltz2018revmodphys,XIONG2022FundamentalResearch}. Therefore, the investigation of these materials paves the way towards not only the fundamental understanding of 2D magnetism, but also the possibility of high-speed and compact AFM spintronic devices.

Standard methods such as magnetization measurements and neutron diffraction, which could only access macroscopic magnetic properties, are not suitable for the study of magnetic structures of atomically thin magnets~\cite{Du2016acsnano}. Especially, antiferromagnets do not have net magnetization, direct measurement of AFM ordering using magneto-optical Kerr effect is not available either. Although recent studies have focused on the Raman spectroscopy~\cite{Du2016acsnano,Kuo2016SciRep,Lee2016Nanolett,Wang2016_2Dmaterials,McCreary2020PRB,Liu2020,Sun2019,Kim2019_2DMater} and the second-harmonic generation~\cite{Chu2020PRL,Ni2021PRL,Sun2019Nature,Ni2021NatNanotech} to detect the crystal symmetry lowering associated with the AFM transition, these signals often do not provide clear identification in the monolayer limit.
Therefore, a comprehensive method which suits for exploring the magnetic properties of 2D antiferromagnets is highly desired.

\begin{figure}[t]
    \centering
    \includegraphics[keepaspectratio,scale=0.2]{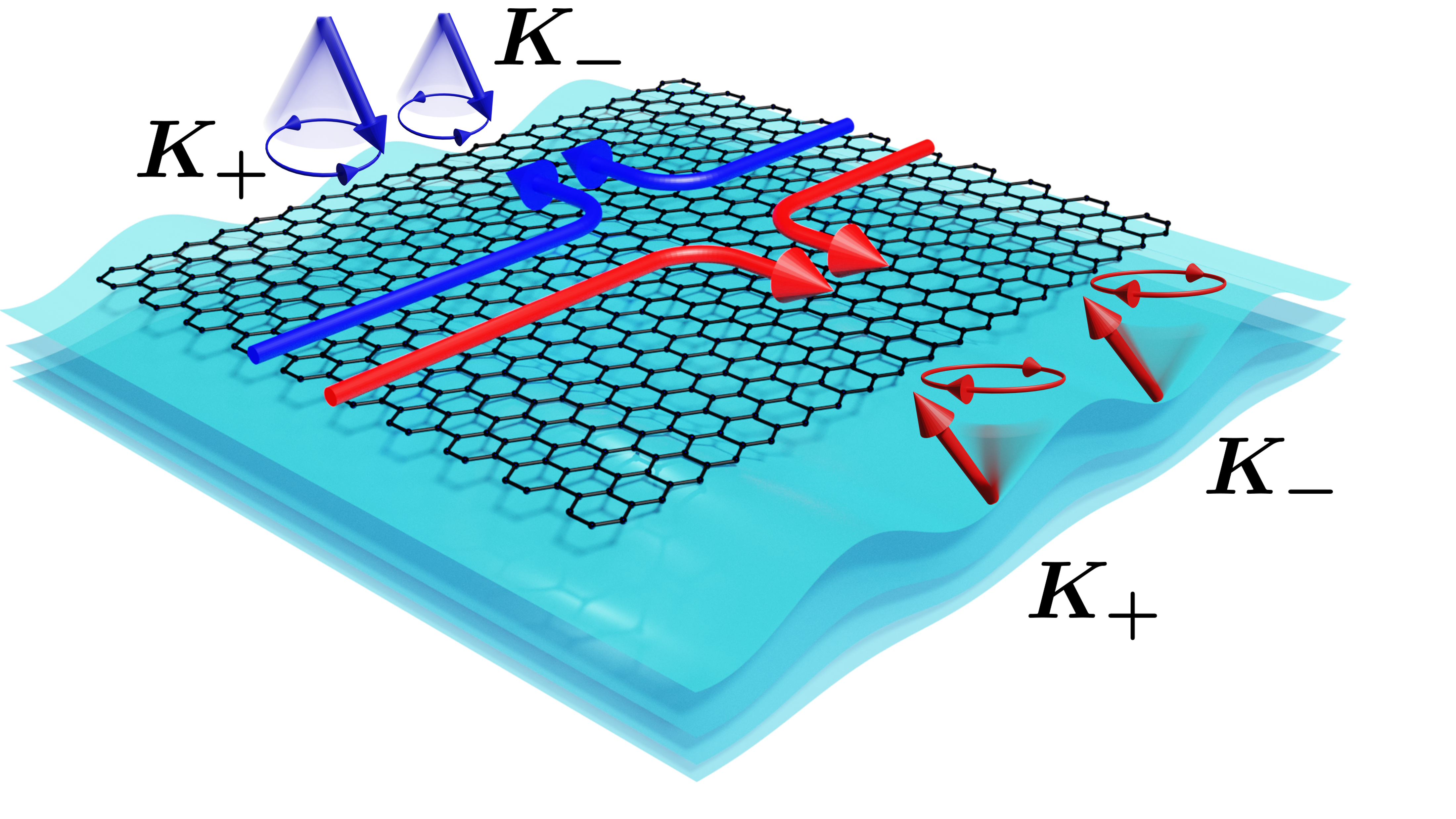}
    \caption{Schematics of acousto-magnonic spin Hall effect. A spatial modulation of the exchange energies due to strain mimics the role of artificial gauge fields for magnons in a honeycomb lattice. Both the strain-induced electric fields and the magnon Berry curvature work at the two valley points in the opposite direction respectively, leading to a net spin Hall current.}

    \label{fig:ARIA}
\end{figure}

Here, we propose an AC intrinsic magnon spin Hall current mechanically driven by surface-acoustic waves (SAWs) as a novel method to probe the magnetic structures of such 2D honeycomb antiferromagnets [see Fig.~\ref{fig:ARIA}]. Owing to extremely large mechanical flexibility of 2D materials, SAWs are ideally suited for fundamental research of them~\cite{Nie2023NanoHoriz}. When an inhomogeneous strain is applied to 2D honeycomb magnets, a spatial modulation of exchange energies mimics the role of artificial gauge fields for magnons~\cite{Ferreiros2018PRB,Nayga2019PRL,Liu2021PRB,Sun2021PRB,Sun2021PRR}. These strain gauge fields work at the two valley points in the opposite direction, which in turn activates the valley degrees of freedom (DOF). Therefore, the valley DOF with the use of SAWs is a promising concept for detection and manipulation of the magnetic order in 2D vdW antiferromagnets. Furthermore, our mechanically driven magnon spin Hall effect will overcome the difficulties with weak magnetic responses inherent in the use of antiferromagnets and hence provide a building block for more sophisticated AFM spintronics.

{\it Formulation.---}
We start from a standard nearest-neighbor AFM Heisenberg model of spins $S$ placed on the sites of a honeycomb lattice:
\begin{equation}
    \hat{H}_0=J\sum_{\langle i,j\rangle}\hat{\vb*{S}}_i\vdot\hat{\vb*{S}}_j,\label{main:Heisenberg}
\end{equation}
where $J>0$ is the AFM exchange interaction, $\hat{\vb*{S}}_i=(\hat{S}^x_i,\hat{S}^y_i,\hat{S}^z_i)$ are the spin operators at the $i$-th site, and the sum $\langle i,j\rangle$ runs over nearest-neighbor sites. Here, we have neglected the magnetic anisotropy because this entails only quantitative changes in the main results~\footnote{See the Supplemental Materials for the effect of the magnetic anisotropy}. We also assume that the ground state of the unstrained Hamiltonian $\hat{H}_0$ is the N\'{e}el ordered state perpendicular to the hexagonal plane and it is maintained under weak strain.

\begin{figure}[t]
    \centering
    \includegraphics[keepaspectratio,scale=0.35]{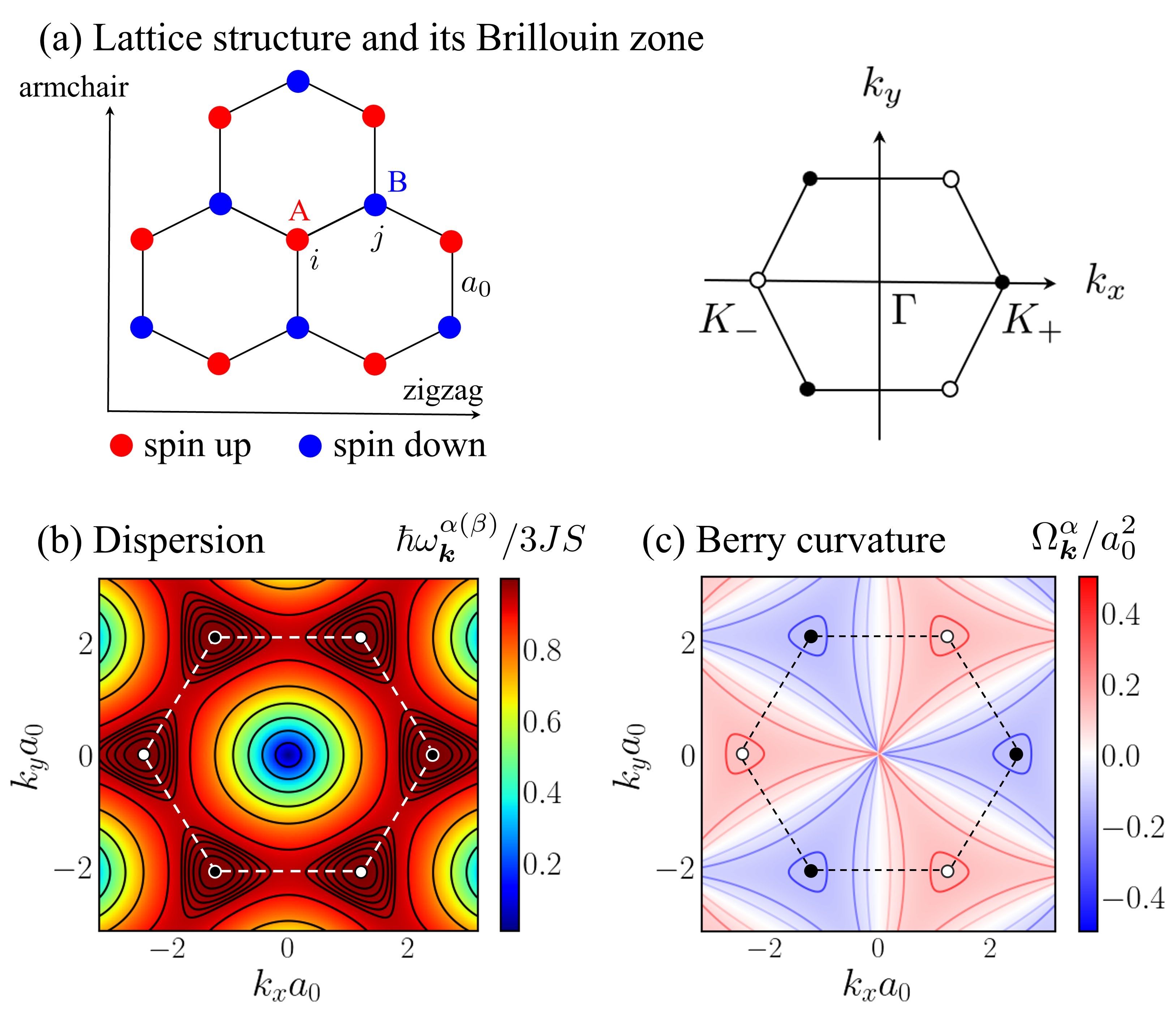}
    \caption{(a) Lattice structure of the N\'{e}el ordered honeycomb antiferromagnets and the corresponding Brillouin zone. Two valley points $K_\pm$ reside at the corners of the hexagonal Brillouin zone. (b) Dispersion of magnons in the unstrained honeycomb Heisenberg antiferromagnets, which shows quadratic maximum at the valley points. (c) Distribution of the magnon Berry curvature for $\alpha$-magnons.}
    \label{fig:dispersion}
\end{figure}

The quantized spin-wave excitations in magnets so-called magnons have attracted special attention as a promising candidate for a spin information carrier with good coherence and without dissipation of the Joule heating~\cite{Uchida2010APL,Kruglyak_2010,Serga_2010,Kajiwara2010,Chumak2015NatPhys,Otani2017,Chumak2017JPhysD,Althammer_2018,Cornelissen2015Natphys,GoennenweinAPL2015,Demidov2017NatCommun,Olsson2020PRX,Schlitz2021PRL}. The excitation spectrum of magnons is obtained by the linear spin-wave theory. We perform the Holstein-Primakoff transformation for magnons on sublattices A and B respectively~\cite{HP1940PhysRev},
\begin{subequations}
    \begin{gather}
    \hat{S}_{i,\mathrm{A}}^+=\sqrt{2S-\hat{a}_i^\dagger\hat{a}_i}\hat{a}_i,\quad\hat{S}_{i,\mathrm{A}}^z=S-\hat{a}_i^\dagger\hat{a}_i,\\
    \hat{S}_{j,\mathrm{B}}^+=\hat{b}_j^\dagger\sqrt{2S-\hat{b}_j^\dagger\hat{b}_j},\quad\hat{S}_{j,\mathrm{B}}^z=-S+\hat{b}_j^\dagger\hat{b}_j,
\end{gather}
\end{subequations}
which describe fluctuations above the N\'{e}el ordered ground state and $\hat{S}^\pm_j=\hat{S}^x_j\pm i\hat{S}^y_j$ are the raising and lowering operators for the $j$-th spin. Here, the bosonic operator $\hat{a}_i$ ($\hat{b}_j^\dagger$) annihilates (creates) a magnon at the $i$ ($j$)-th A (B) site. The Hamiltonian is then diagonalized by subsequent Fourier and Bogoliubov transformations: $\hat{\alpha}_{\vb*{k}}=u_{\vb*{k}}\hat{a}_{\vb*{k}}-v_{\vb*{k}}\hat{b}_{-\vb*{k}}^\dagger$ and $\hat{\beta}^\dagger_{-\vb*{k}}=u_{\vb*{k}}\hat{b}_{-\vb*{k}}^\dagger-v_{\vb*{k}}^\ast\hat{a}_{\vb*{k}}$, and Eq.~\eqref{main:Heisenberg} is cast into non-interacting Dirac magnons~\cite{Nayga2019PRL},
\begin{equation}
    \hat{H}_0=\sum_{\vb*{k}}(\hbar\omega_{\vb*{k}}^\alpha\hat{\alpha}^\dagger_{\vb*{k}}\hat{\alpha}_{\vb*{k}}+\hbar\omega_{\vb*{k}}^\beta\hat{\beta}^\dagger_{\vb*{k}}\hat{\beta}_{\vb*{k}}),\label{main:hamiltonian}
\end{equation}
which is justified well below the N\'{e}el temperature. In these honeycomb systems, two inequivalent valley points $K_\pm$ reside at the corners of the hexagonal Brillouin zone [see Fig.~\ref{fig:dispersion}]. In the vicinity of $K_\pm$, $\hat{\alpha}_{\vb*{k}}$ $(\hat{\beta}_{\vb*{k}})$ can be regarded as $\hat{a}_{\vb*{k}}$ $(\hat{b}_{\vb*{k}})$ because the Bogoliubov coefficients are approximated as $u_{K_\pm}\simeq1$ and $v_{K_\pm}\simeq0$. The relevant spectrum of magnons near the valley points are described by a quadratic dispersion: $\hbar\omega^{\alpha(\beta)}_{K_\pm}(\vb*{q})=3JS\sqrt{1-a_0^2\vb*{q}^2/4}$ with relative momentum $\vb*{q}=(q_x,q_y)$ measured from the valley center. We also obtain the magnon Berry curvature $\vb*{\Omega}^{\alpha}_{\vb*{k}}=-\vb*{\Omega}^\beta_{\vb*{k}}$ and its distribution is depicted in Fig.~\ref{fig:dispersion}(c). Notice that $\vb*{\Omega}^\alpha_{\vb*{k}}$ is an odd function and it shows opposite values in the vicinity of $K_\pm$.

A continuum model for Dirac magnons is complemented with elasticity theory~\cite{landau1986elasticity} to incorporate the effect of strain. We here consider a spatial modulation of the exchange energies due to strain which mimics the role of artificial gauge fields that govern magnon dynamics. For large lattice, the displacement field $\vb*{u}(\vb*{r},t)=(u_x,u_y,u_z)$ can be taken as a smooth function of the coordinates, and the strain tensor is defined by $\varepsilon_{ij}=(\partial_iu_j+\partial_ju_i+\partial_ih\partial_jh)/2$ having $h=u_z$ as the normal component of the displacement. By expanding the operators around the two valley points: $\hat{a}(\vb*{r})\simeq e^{i\vb*{K}_\pm\vdot\vb*{r}}\hat{\psi}_a^{K_\pm}(\vb*{r})$, $\hat{b}(\vb*{r})\simeq e^{-i\vb*{K}_\pm\vdot\vb*{r}}\hat{\psi}_b^{K_\pm}(\vb*{r})$, we obtain the strained Hamiltonian as~\footnote{See the Supplemental Materials for the detailed derivation of the effective Hamiltonian under strain.}
\begin{equation}
\hat{H}=\int\dd{\vb*{r}}\hat{\vb*{\Psi}}^\dagger(\vb*{r})\left(
    \begin{array}{cc}
       \mathcal{H}^{K_+}  & 0 \\
       0  & \mathcal{H}^{K_-}
    \end{array}
    \right)\hat{\vb*{\Psi}}(\vb*{r}),
\end{equation}
where $\hat{\vb*{\Psi}}^\dagger=(\hat{\psi}_a^{K_+\dagger},\hat{\psi}_b^{K_+},\hat{\psi}_a^{K_-\dagger},\hat{\psi}_b^{K_-})$ and the continuous effective Hamiltonian is given by
\begin{equation}
    \mathcal{H}^{K_\eta}=-v_J(-i\hbar\grad+\eta\vb*{A}^s)\vdot\vb*{\sigma}_\eta+3JS-\phi^s.
\end{equation}
$v_J=3JSa_0/2\hbar$ is the exchange velocity of magnons with the length of the bond between neighboring magnetic atoms $a_0$. Here, $\eta=\pm1$ is the valley index labelling the two inequivalent valleys $K_\pm$ and $\vb*{\sigma}_\eta=(\eta\sigma_x,\sigma_y)$ are the Pauli matrices representing a psuedospin from the sublattice DOF. The vector potential $\vb*{A}^s=\hbar\frac{\gamma}{2a_0}(\varepsilon_{xx}-\varepsilon_{yy},-2\varepsilon_{xy})$ couples minimally to the magnonic excitations with opposite signs at the two valley points, which results in the activation of the valley DOF. The scalar potential $\phi^\mathrm{s}=\frac{3JS}{2}\gamma(\varepsilon_{xx}+\varepsilon_{yy})$ originates from the change of density due to the variation of the sample area described by $\div\vb*{u}$. The factor $\gamma$ encodes the strength of the magneto-elastic coupling, which is expected to be of the order of unity~\cite{Levy2010Science}. Note that the emergence of the strain gauge fields is related to the fact that quasiparticles in Dirac systems are described by the corresponding relativisticlike equations~\cite{Vozmediano2010PhysRep,Juan2013PRB,Naumis2017RepProgPhys,Zhai2019ModPhysLettB,Si-YuLi2020PRL}. 

The introduction of the gauge fields ($\phi^s, \vb*{A}^s$) leads to emergent pseudo electromagnetic fields:
\begin{subequations}\label{main:EB}
    \begin{gather}
    \vb*{E}_1^s=-\grad\phi^s,\quad\vb*{E}_2^s=-\partial_t\vb*{A}^s,\\ \vb*{B}^s=\curl\vb*{A}^s.
    \end{gather}
\end{subequations}
Here, $\vb*{E}_2^s$ is generated only when considering a time-varying strain such as SAWs and has a significant importance for the main results. Fundamentally, this pseudo electric field couples not the charge DOF but the valley, and hence opens a new paradigm to explore novel properties of charge-neutral quasiparticles~\cite{Massarelli2017PRB,Nica2018PRB,Rachel2016PRL}. Previous magnon transport in conventional antiferromagnets relies on thermal gradients as a driving force~\cite{Cheng2016PRL,Zyuzin2016PRL,Shiomi2017PRB,Takashima2018PRB,Kondo2022PRR,Feringa2023PRB}, but these alone cannot distinguish the spin between two magnons; therefore, a magnon-mediated spin current has been difficult to generate without lifting the degeneracy of two magnons. However, in honeycomb antiferromagnets, the valley DOF is activeted by the strain gauge fields and hence has a potential to result in a pure spin current by combining with the valley-contrasting magnon Berry curvature.

The semiclassical equations for magnons under strain-induced pseudo electromagnetic fields Eq.~\eqref{main:EB} are derived in a similar form to the conventional semiclassical equations for electrons%~\cite{DiXiao2010review}
~\cite{DiXiao2010review},
\begin{subequations}\label{main:Semi}
\begin{gather}
\dot{\vb*{r}}^{\alpha(\beta)}_{\eta}=\pdv{\omega^{\alpha(\beta)}_{\eta}}{\vb*{q}}-\dot{\vb*{q}}^{\alpha(\beta)}_\eta\times\vb*{\Omega}_{\eta}^{\alpha(\beta)},\\
\hbar\dot{\vb*{q}}^{\alpha(\beta)}_\eta=-\vb*{{E}}^s_\eta-\dot{\vb*{r}}^{\alpha(\beta)}_\eta\times\vb*{{B}}^s_\eta,
\end{gather}
\end{subequations}
where we have introduced a compact form of pseudo-electromagnetic fields: $\vb*{B}^s_\eta=\eta\vb*{B}^\mathrm{s}$, $\vb*{E}_\eta^s=\vb*{E}^s_1+\eta\vb*{E}_2^s$. We should note the validity of the treatment in the framework of the semiclassical approach. The typical frequencies of SAWs range from MHz to GHz, whereas those of relevant magnons in the vicinity of valley points are of the order of THz; therefore, we can assume that magnons adiabatically follow the deformation and their dynamics are governed by Eq.~\eqref{main:Semi}.

\begin{figure}
    \centering
    \includegraphics[keepaspectratio,scale=0.45]{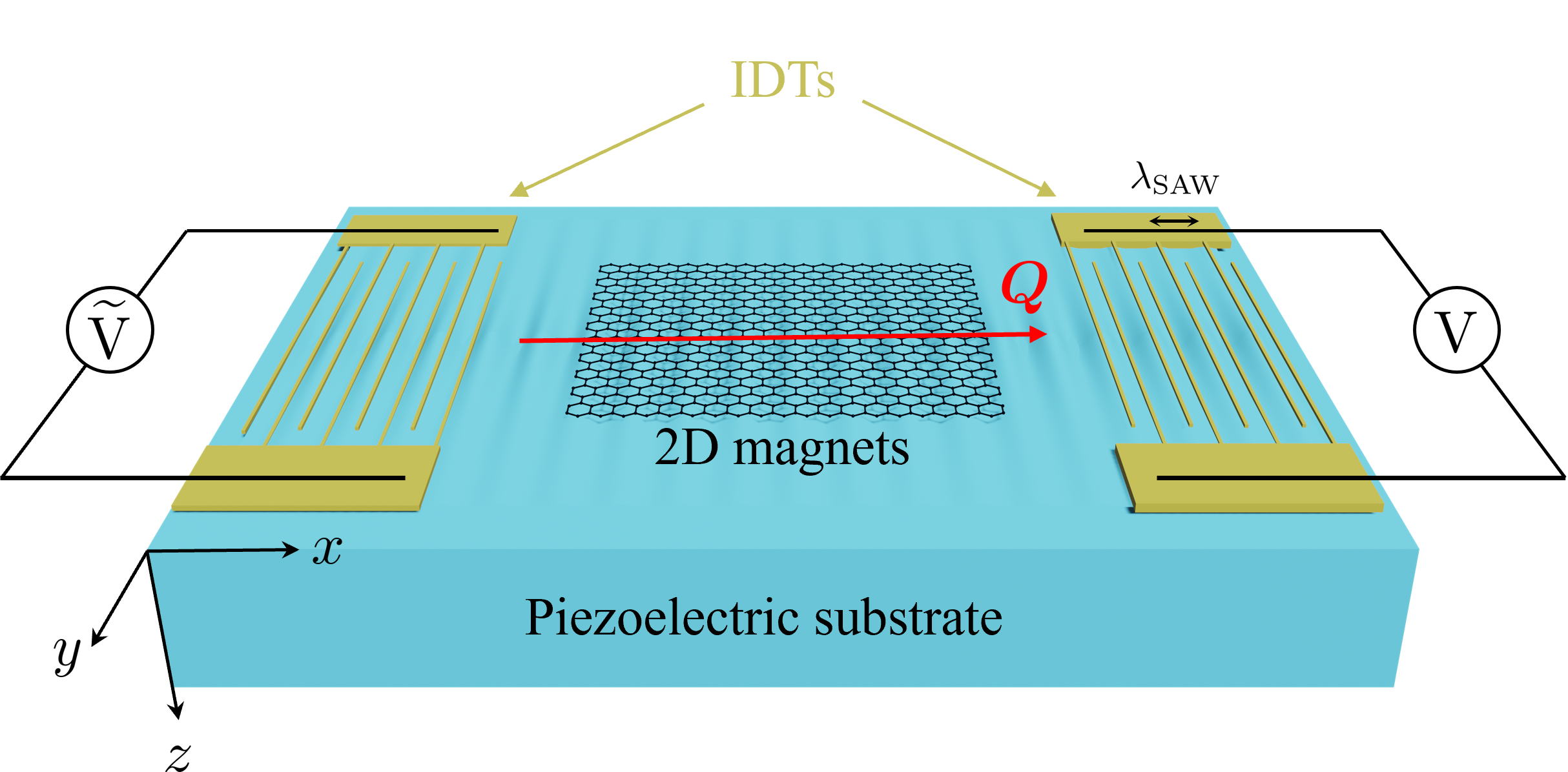}
    \caption{Schematic illustration of the Rayleigh-type SAWs generation. SAWs are mainly generated, detected, and controlled by interdigital transducers (IDTs) placed on the ends of a piezoelectric substrate. IDTs are periodic arrays of metallic finger electrodes with a pitch of half the SAW wavelength $\lambda_\mathrm{SAW}=2\pi/Q$ and convert AC electric signals into the SAWs % with frequency $\omega$ 
    propagating along $\vb*{Q}$ via the inverse piezoelectric effect.}
    \label{fig:SAW-IDT}
\end{figure}

{\it Rayleigh-type SAWs-induced pseudo electric fields.---}
Among the diverse modes of SAWs, the Rayleigh-type waves, which are the superposition of longitudinal and normal components, can be easily excited under traction-free boundary conditions on piezoelectric substrates~\cite{Nie2023NanoHoriz}. Considering Rayleigh-type SAWs propagating on the surface of a piezoelectric substrate in the $xy$-plane [see Fig.~\ref{fig:SAW-IDT}], the displacement field is given by
\begin{equation}
    \vb*{u}_{\mathrm{Rayleigh}}(\vb*{r},t)=\Re[(u_L\hat{\vb* {Q}}+iu_z\hat{\vb*{z}})e^{i(\vb*{Q}\vdot\vb*{r}-\omega t)}],
\end{equation}
where $u_L$ and $u_z$ are the longitudinal and normal displacements, $\vb*{Q}=Q(\cos\theta,\sin\theta)$ is the in-plane propagating wave vector with $\theta$ being an azimuthal angle, and $\omega$ is the frequency of applying SAWs. Here, $\theta=0$ corresponds to the $x$-direction with the zigzag orientation of the honeycomb lattice. By assuming that the van der Waals magnet on a piezoelectric substrate completely follows the displacement of the substrate, the Rayleigh-type SAWs-induced pseudo electric fields reads
\begin{align}\label{main:gauge}
    \vb*{E}_2^s=\hbar\frac{\gamma}{2a_0}c_t\xi Q^2\Re[u_L(-\cos3\theta\hat{\vb*{Q}}+\sin3\theta\hat{\vb*{\theta}})e^{i(\vb*{Q}\vdot\vb*{r}-\omega t)}],
\end{align}
where $\hat{\vb*{\theta}}=\partial_\theta\hat{\vb*{Q}}=(-\sin\theta,\cos\theta)$ is the azimuthal unit vector transverse to $\hat{\vb*{Q}}$, $c_t$ is the transverse velocity of the sound wave, and $\xi$ is a constant characterizing the SAWs dispersion as $\omega=c_t\xi Q$. We should note that the pseudo electromagnetic fields stemming from the out-of-plane displacement are proportional to $u_z^2$ due to $\partial_ih\partial_jh$, which is less relevant under weak strain, and hence we neglect their contributions in the following analysis.

{\it Acousto-magnonic spin Hall effect.---}
As a demonstration of semiclassical magnon transport driven by SAWs-induced gauge fields, we consider magnon-mediated spin currents. Because the two valley points are largely separated in the Brillouin zone, the valley index $\eta$ can be used as a good quantum number in the presence of weak disorder. 
In previous works, various electron transport phenomena under strain applied to 2D Dirac materials have been intensively studied~\cite{Oppen2009PRB,Miseikis2012APL,Vaezi2013PRB,Kalameitsev2019PRL,Sonowal2020PRB,Sukhachov2020PRL,Sela2020PRL,Bhalla2022PRB,Zhao2022PRL,Ominato2022PRB}. However, the role of SAWs in magnon transport has not been well investigated yet.

Since the two magnons are completely decoupled in Eq.~\eqref{main:hamiltonian}, we can treat the dynamics of each mode independently. The Boltzmann equation for $\alpha$($\beta$)-magnons is then given by,
\begin{equation}
\pdv{n^{\alpha(\beta)}_{\eta}}{t}+\dot{\vb*{r}}_{\eta}\vdot\pdv{n^{\alpha(\beta)}_{\eta}}{\vb*{r}}+\dot{\vb*{q}}_{\eta}\vdot\pdv{n^{\alpha(\beta)}_{\eta}}{\vb*{q}}=-\frac{n^{\alpha(\beta)}_{\eta}-%n^{\mathrm{eq}}
n_\mathrm{B}(\hbar\omega_{\vb*{q}})}{\tau},
\end{equation}
where $n^{\alpha(\beta)}_{\eta}(\vb*{r},\vb*{q},t)$ %and $n^{\mathrm{eq}}=n_\mathrm{B}(\hbar\omega_{\vb*{q}})$ are the local and the global equilibrium 
is the distribution function for $\alpha$ ($\beta$)-magnons with valley $\eta$ and relative momentum $\vb*{q}$. $n_\mathrm{B}(\epsilon)=(e^{\epsilon/k_\mathrm{B}T}-1)^{-1}$ is the Bose-Einstein distribution function with zero chemical potential and $\tau$ is the momentum relaxation time for magnons.

We are now ready to discuss the magnon-mediated spin currents driven by the Rayleigh-type SAWs. By invoking the Bogoliubov transformation, we obtain the $z$ component of the total spin as $\hbar\sum_{\vb*{k}}(-\hat{\alpha}_{\vb*{k}}^\dagger \hat{\alpha}_{\vb*{k}}+\hat{\beta}_{\vb*{k}}^\dagger \hat{\beta}_{\vb*{k}})$, and thereby $-\hbar$ ($+\hbar$) spin angular momentum is carried by $\alpha$ ($\beta$)-magnons~\cite{Rezende2019JApplPhys}. Because the strain gauge fields only work around the two valley points and magnons in other points do not contribute to the spin currents due to their degeneracy~\footnote{See the Supplemental Materials for detailed calculations.}, we only consider the contribution from the vicinity of $K_\pm$:
\begin{equation}
    \vb*{j}_{\mathrm{s}}^z=\hbar\sum_{\eta=\pm1}\int[\dd{\vb*{q}}]\left(-\mathcal{D}^\alpha\dot{\vb*{r}}_\eta^{\alpha} n^{\alpha}_{\eta}+\mathcal{D}^\beta\dot{\vb*{r}}_\eta^{\beta} n^{\beta}_{\eta}\right),\label{main:spincurrent}
\end{equation}
where the factor $\mathcal{D}^{\alpha(\beta)}\equiv1+\frac{1}{\hbar}\vb*{B}_\eta^s\vdot\vb*{\Omega}^{\alpha(\beta)}_{\eta}(\vb*{q})$ is a field-induced correction to the volume of the phase space and $\int[\dd{\vb*{q}}]\equiv\int\dd[2]{\vb*{q}}/(2\pi)^2$. By substituting Eq.~\eqref{main:Semi} into Eq.~\eqref{main:spincurrent}, we obtain a transverse magnon-mediated spin current as
\begin{align}
    \vb*{j}_s^z&=\vb*{E}_2^s\times\int[\dd{\vb*{q}}](-\vb*{\Omega}^\alpha_+ n^\alpha_++\vb*{\Omega}^\alpha_-n^\alpha_-+\vb*{\Omega}^\beta_+ n^\beta_+-\vb*{\Omega}^\beta_-n^\beta_-)\nonumber\\
    &=-\vb*{E}_2^s\times\int[\dd{\vb*{q}}]\vb*{\Omega}^\alpha_{+}(n^\alpha_{+}+n^\alpha_{-}+n^\beta_{+}+n^\beta_{-}),\label{main:AMSHE}
\end{align}
where we have used the relation $\vb*{\Omega}_{\vb*{k}}^{\alpha}=-\vb*{\Omega}^\alpha_{-\vb*{k}}=-\vb*{\Omega}^\beta_{\vb*{k}}$~\footnote{Our strategy provides an intrinsic magnon spin Hall effect even in topologically trivial antiferromagnets: $\int[\dd{\vb*{k}}]\vb*{\Omega}^{\alpha(\beta)}=0$. This fact is quite different from the previous studies on the magnon thermal Hall effect, which relies on topologically nontrivial magnon bands. Furthermore, the thermal Hall current vanishes in the absence of the Dzyaloshinskii-Moriya interaction, and hence the only remaining net current is the spin Hall current. For detailed calculations, see the Supplemental Materials.}. Equation~\eqref{main:AMSHE} is the main result of this Letter, which originates from the interplay between the strain gauge fields and the magnon Berry curvature. Both of them have a valley-contrasting property and work in the opposite direction at the two valley points respectively, resulting in a net spin Hall current [see Fig.~\ref{fig:ARIA}]. Furthermore, Eq.~\eqref{main:AMSHE} gives a finite contribution even when the magnons obey the Bose-Einstein distribution $n_\mathrm{B}$; therefore, our mechanically driven spin Hall current dubbed acousto-magnonic spin Hall effect becomes independent of the magnon relaxation time $\tau$ and hence provides an intrinsic spin Hall current.

\begin{figure}[t]
    \centering
    \includegraphics[keepaspectratio,scale=0.3]{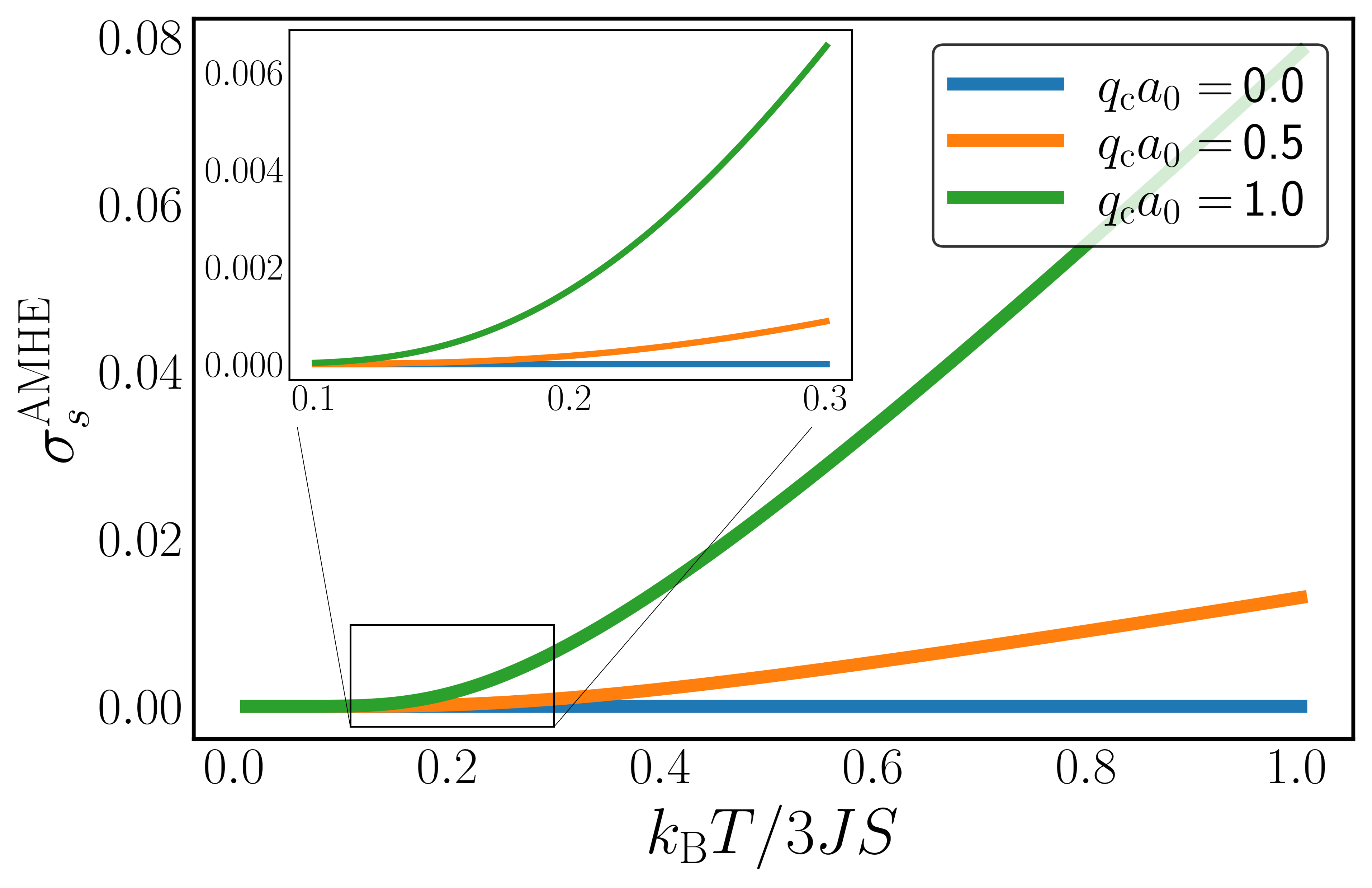}
    \caption{Temperature dependence of the acousto-magnonic spin Hall conductivity $\sigma_s^{\mathrm{AMHE}}$ for different values of $q_ca_0$. The horizontal axis is normalized by $3JS/k_\mathrm{B}$.}
    \label{fig:spin_current}
\end{figure}

{\it Discussion.---}%
Finally, we will discuss the experimental feasibility of our acousto-magnonic spin Hall effect. %The experimental realization can be achieved with a van der Waals layered magnet which is placed on a piezoelectric substrate. 
Transition metal phosphorus trichalcogenides $M$P$X_3$ are a family of AFM semiconductors with a bandgap of the order of 1\,eV~\cite{Och2021Nanoscale,Jiang2021APR,Xu2022Microstructures,Wang2022ACSNano,Kurebayashi2022NatRevPhys}, which is much larger than the typical frequency of SAWs, and hence the spin transport is dominated by magnons. Here, we suppose MnPS$_3$ as a candidate of N\'{e}el-type antiferromagnets~\cite{Sun2019,Kim2019_2DMater,Chu2020PRL,Ni2021PRL}, in which long-distance magnon transport over several micrometers has been recently observed~\cite{Xing2019PRX}. By introducing the cut-off wave number $q_c$, which is a radius of the effective region of strain gauge fields in the vicinity of each valley point, the spin Hall conductivity is approximated as
\begin{align}
    \sigma_s^{\mathrm{AMHE}}&=\int_{3JS\sqrt{1-a_0^2q_c^2/4}}^{3JS}\dd{\epsilon}D(\epsilon)\Omega^\alpha(\epsilon) n_\mathrm{B}(\epsilon)\nonumber\\
    &\simeq\frac{q_c^2a_0^2}{4\pi}\frac{1}{e^{3JS/k_\mathrm{B}T}-1},
\end{align}
where $D(\epsilon)$ is the density of states and we have defined the spin Hall conductivity owing to acousto-magnonic spin Hall effect as $\vb*{j}_s^z=\sigma_s^{\mathrm{AMHE}}\vb*{E}_2^s\times\hat{\vb*{z}}$. Fig.~\ref{fig:spin_current} shows the temperature and $q_c$ dependences of $\sigma_s^{\mathrm{AMHE}}$, in which the horizontal axis is normalized by $3JS/k_\mathrm{B}\simeq$\,134 K. The N\'{e}el temperature is experimentally obtained as $T_\mathrm{N}\simeq79$\,K~\cite{Xing2019PRX,Ni2021PRL}. Therefore, the amplitude of the spin current $\abs{\vb*{j}_s^z}$ is estimated to be of the order of $\mathrm{meV/m}$ at $T\ll T_\mathrm{N}$ with the pseudo electric field $\abs{\vb*{E}_2^s}\simeq1\,\mathrm{eV/m}$ induced by the Rayleigh-type SAWs. Here, we have used the parameters~\cite{Xing2019PRX}: $J\simeq1.54$\,meV, $S=5/2$, $a_0\sim1$\,\AA, $u_L\simeq100$\,pm, $\lambda_\mathrm{SAW}=2\pi/Q\simeq1$\,$\mu$m, $c_t=4000\,\mathrm{m/s}$, $\xi\simeq0.95$, and $\gamma\sim1$~\cite{Levy2010Science}. This AC spin current oscillating in the order of GHz may be detected by the present experiments such as spin wave resonance~\cite{Kobayashi2017PRL,Tateno2020PRB}.

The conventional strategy for generating magnon spin currents in AMF spintronics has been applying magnetic fields in order to lift the degeneracy of magnons or to realize a canted AFM structure along the fields~\cite{Jungwirth2016NatNanotech,Baltz2018revmodphys,XIONG2022FundamentalResearch}. This provides only weak spin currents due to extrinsically perturbative magnetic fields. On the other hand, the valley DOF, which is a characteristic of the hoenycomb lattice structure, is activated by the strain-induced gauge fields. Here, we have focused on the interplay between the strain gauge fields and the magnon Berry curvature, which have opposite sign between the two valley points respectively, resulting in an intrinsic magnon spin Hall current. This fact may offer a bright prospect for the long-standing dilemma %for null stray field and robustness against extenal magnetic fields 
that antiferromagnets show only weak magnetic responses which are hard to detect and control. Furthermore, our acousto-magnonic spin Hall effect does not require the Dzyaloshinskii-Moriya interaction, which is an essential ingredient for the magnon spin Nernst effect~\cite{Cheng2016PRL,Zyuzin2016PRL,Shiomi2017PRB} but is negligible in MnPS$_3$~\cite{WildesPRB2021}. In addition, our results do not rely on the nontrivial topology of magnon bands originates from the magnon-phonon coupling. However, the magnon-phonon coupling results in the formation of the hybridized excitation of magnons and phonons, which can carry large Berry curvature in the anticrossing regions between their bands~\cite{Thingstad2019PRL,Gyungchoon2019PRL,Park2020NanoLett,Zhang2020PRL,Bazazzadeh2021PRB}. The impact of the magnon-phonon coupling on our results may be an interesting future work.
Therefore, our results will overcome the difficulties inherent in the use of antiferromagnets and provide a building block for more sophisticated AFM spintronics.

\begin{table}[t]
     \centering
    \caption{Symmetry classification and comparison of the conventional thermal gradient and the strain-induced pseudo-electric fields introduced in this work under the inversion $\mathcal{P}$, the time reversal $\mathcal{T}$, and the effective time reversal $\mathcal{TC}$ symmetry operations.}
    \begin{tabular}{cc|ccc}
     \hline
     \multicolumn{2}{c|}{Input/Output fields} & $\mathcal{P}$ & $\mathcal{T}$ & $\mathcal{TC}$\\
     \hline \hline
     Thermal gradient & $\grad T$ & $-$ & $+$& $+$\\
     \hline
     Strain tensor & $\varepsilon_{ij}$ & $+$ & $+$ & $+$\\
     Strain gauge fields & $\phi^s$, $\vb*{A}^s$ & $+$ & $+$ & $+$\\
     Pseudo-electric field & $\vb*{E}_2^s$ & $+$ & $-$ & $-$\\
     \hline
     Magnon spin current & $\vb*{j}_s^z$ & $-$ & $+$ & $-$\\
     \hline 
     \end{tabular}
     \label{tab:symmetry}
\end{table}
 
Symmetry considerations are summarized in Table~\ref{tab:symmetry}. We can see that the symmetry of $\vb*{E}_2^s$ is quite different from that of the conventional thermal gradient. Thus, $\vb*{E}_2^s$-induced magnon spin currents are prohibited in centrosymmetric antiferromagnets and our acousto-magnonic spin Hall effect becomes a novel probe for noncentrosymmetric AFM phases such as the N\'{e}el ordered phase. Furthermore, $\vb*{E}_2^s$ can generate magnon spin current with preserving the effective time reversal symmetry $\mathcal{TC}$, which is the combined action of time reversal $\mathcal{T}$ and $180^\circ$ spin rotation around an in-plane axis $\mathcal{C}$. This is why our acousto-magnonic spin Hall effect does not require the Dzyaloshinskii-Moriya interaction and the magnon-phonon coupling. 

{\it Conclusion.---}%
In summary, we have developed a basic framework of SAWs-driven magnon transport in a honeycomb antiferromagnet. Here, we have proposed a magnon-mediated spin Hall current driven by the Rayleigh-type SAWs dubbed acousto-magnonic spin Hall effect as a novel probe for exploring the magnetic properties of such 2D vdW antiferromagnets. By focusing on the valley DOF, we have revealed that the interplay between the strain gauge fields and the magnon Berry curvature results in an intrinsic spin Hall current without the Dyzaloshinskii-Moriya interaction and the magnon-phonon coupling in the N\'{e}el ordered state. Therefore, our results open a promising route for mechanical detection and manipulation of the magnetic order in 2D antiferromagnets. Furthermore, they will overcome the difficulties with weak magnetic responses inherent in the use of antiferromagnets and hence provide a building block for future AFM spintronics. Recent studies have shown that the strain effect also introduce the pseudo-gauge fields for Dirac magnons in honeycomb ferromagnets~\cite{Ferreiros2018PRB}, twisted honeycomb ferromagnet~\cite{Liu2021PRB}, and honeycomb noncollinear antiferromagnets~\cite{Owerre2017JPhysCommun}. Therefore, magnon Hall effects driven by the surface acoustic waves may be applicable to a wide range of 2D magnets and are our interesting future work. Furthermore, we expect that the measurements of the magnon Hall effect based on our strategy is robust against electronic contributions~\footnote{The electronic spin Hall effect is prohibited under the time-reversal symmetry as can be seen in the Table~\ref{tab:symmetry}. Therefore, the emergence of a net spin Hall current by pseudo electric fields is unique to the magnonic systems.}. Our work motivates further systematic studies on 2D vdW magnets, which are significant not only for the potentially diverse applications, but also for the fundamental understanding of 2D magnetism.

\begin{acknowledgements}
The authors are grateful to K. Shinada, M. Chazono, S. Kaneshiro, T. Funato, J. Fujimoto, and D. Oue for valuable discussions. R.S. thanks Y. Nozaki, K. Yamanoi, T. Horaguchi, S. Watanabe, and S. Fujii for providing helpful comments from an experimental point of view. We also thank the referees for noticing the Refs.~\cite{Thingstad2019PRL,Gyungchoon2019PRL,Park2020NanoLett,Zhang2020PRL,Bazazzadeh2021PRB}. R.S. is supported by JSPS KAKENHI (Grants JP 22J20221 and 22KJ1937). This work was supported by the Priority Program of Chinese Academy of Sciences under Grant No. XDB28000000, and by JSPS KAKENHI for Grants (No. 20H01863, 21H04565, and 23H01839) from MEXT, Japan.
\end{acknowledgements}

\bibliography{ref}

\ifarXiv
    \foreach \x in {1,...,\numbersupplementpages}
    {
        \clearpage
        \includepdf[pages={\x,{}}]{\supplementfilename}
    }
\fi

\end{document}